\pdfoutput=1
\RequirePackage{ifpdf}
\ifpdf 
\documentclass[pdftex]{sigma}
\else
\documentclass{sigma}
\fi

\newtheorem{Theorem}{Theorem}
\newtheorem{Proposition}[Theorem]{Proposition}
\newtheorem{Conjecture}[Theorem]{Conjecture}
\newtheorem{question}[Theorem]{Question}

\DeclareMathOperator{\OB}{OB}
\DeclareMathOperator{\PB}{PB}

\begin{document}

\newcommand{\arXivNumber}{1210.2812}

\allowdisplaybreaks

\renewcommand{\PaperNumber}{095}

\FirstPageHeading

\ShortArticleName{Algebraic Geometry of Matrix Product States}

\ArticleName{Algebraic Geometry of Matrix Product States}

\Author{Andrew CRITCH~$^\dag$ and Jason MORTON~$^\ddag$}

\AuthorNameForHeading{A.~Critch and J.~Morton}

\Address{$^\dag$~Jane Street Capital, 1 New York Plaza New York, NY 10004, USA}
\EmailD{\href{mailto:critch@acritch.com}{critch@acritch.com}}
\URLaddressD{\url{http://www.acritch.com}}

\Address{$^\ddag$~Department of Mathematics, Pennsylvania State University, University Park, PA 16802, USA}
\EmailD{\href{mailto:morton@math.psu.edu}{morton@math.psu.edu}}
\URLaddressD{\url{http://www.jasonmorton.com}}

\ArticleDates{Received February 28, 2014, in f\/inal form August 22, 2014; Published online September 10, 2014}

\Abstract{We quantify the representational power of matrix product states (MPS) for entangled qubit systems by giving
polynomial expressions in a~pure quantum state's amplitudes which hold if and only if the state is a~translation
invariant matrix product state or a~limit of such states.
For systems with few qubits, we give these equations explicitly, considering both periodic and open boundary conditions.
Using the classical theory of trace varieties and trace algebras, we explain the relationship between MPS and hidden
Markov models and exploit this relationship to derive useful parameterizations of MPS.
We make four conjectures on the identif\/iability of MPS parameters.}


\Classification{14J81; 81Q80; 14Q15}

Matrix product states (MPS) provide a~useful and popular model of 1-D quantum spin systems which approximate the ground
states of gapped local Hamiltonians~\cite{verstraete2006matrix}.
Here we describe two results concerning the algebraic geometry of such models.

First, with periodic or open boundary conditions, we describe the closure of the set of states representable~by
translation invariant binary MPS as an algebraic variety.
Our description is given as an ideal of polynomials in the amplitudes of the state that vanish if and only if the state
is a~limit of MPS with~$N$ spins and $D = d = 2$ dimensional virtual and physical bonds.
In small cases our description is complete.
In general such implicitization problems are very dif\/f\/icult.
In Section~\ref{sec-cmps}, we exhibit a~polynomial which vanishes on a~pure state if and only if it is a~limit of binary
translation invariant, periodic boundary MPS with $N=4$, and a~set of $30$ polynomials which vanish when $N=5$.
We also obtain many linear equations which are satisf\/ied for~$N$ up to~$12$.
In Section~\ref{sec-bmps}, Theorem~\ref{thm-hyp2} gives an analogous result for MPS with open boundary conditions and
$N=3$.
Finally we examine cases where $N\gg 0$.
While related, determining the ideal of the variety of MPS is distinct from problems such as f\/inding entanglement
monotones and invariants under local unitary or local special linear group actions.

Matrix product states bear a~close relationship to probabilistic graphical models known as {\em hidden Markov models}
(HMM)~\cite{EDHMM}.
Our second main result, described in Section~\ref{sec:HMM}, is to make this relationship precise by modifying the
parametrization of HMM to obtain MPS.
We review the invariant theory of trace identities and trace varieties~\cite{procesi1976invariant} that has been used to
study HMM~\cite{critch2013binary}, and how these results apply to varieties of MPS.
In particular we obtain a~nice parametrization for translation invariant binary MPS with periodic boundary conditions.
Such parameterizations, by minimizing redundancy, reduce the dimensionality of the optimization problems arising in the use
of tensor network models to study physical phenomena.

Finally in Section~\ref{sec:conc} we suggest a~``dictionary'' of similar relationships between probabilistic graphical
models and tensor network state models.
Our results are complimentary to the connection between invariant theory and diagrammatic representations explored
in~\cite{biamonte2012invariant}.
In~\cite{morton2014invariant}, the appropriate generalization of the trace algebra~\cite{procesi1976invariant} for
higher dimensional analogues of MPS (such as PEPS) is derived.

\section{Representability by translation invariant matrix\\ product states}

\label{sec-cmps}
\begin{figure}[t]
\centering
\includegraphics{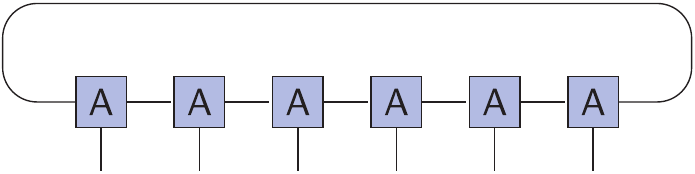}

\caption{Translation-invariant MPS with periodic boundary.}
\label{fig:TIMPSPB}
\end{figure}

First consider a~translation-invariant matrix product state with {\em periodic boundary} conditions (see Fig.~\ref{fig:TIMPSPB}).
Suppose the inner (virtual) bond dimension is~$D$, the outer (physical) bond dimension is~$d$, and there are~$N$ spins.
Fix $D \times D$ complex parameter matrices $A_{0}, \dots, A_{d-1}$, def\/ining the same $D  \times  D  \times d$
parameter tensor at each site.
This def\/ines the tensor network state, for $i_j \in \{0, \dots, d-1\}$,
\begin{gather}
\label{eqn-phi}
\Psi = \sum\limits_{i_1, \dots, i_N} \operatorname{tr}(A_{i_1} \cdots A_{i_N})\,|\,i_1 i_2 \dots i_N \rangle.
\end{gather}
\begin{question}
\label{q:characterize}
Fixing virtual and physical bond dimension, which states are matrix product states?
\end{question}
Including states which are limits of MPS, a~precise answer to this question could be given as a~constructive description
of the set of polynomials~$f$ in the coef\/f\/icients of~$\Psi$ such that $f(\psi_{i_1, \dots, i_N})=0$ if and only if~$\psi$
is a~limit of MPS.
This would describe the (closure of the) set of MPS as an algebraic variety.
See~\cite{Cox1997,Eisenbud2002,Greuel2008} for background on varieties and computational commutative algebra.

Such a~description is possible because of the way MPS are def\/ined.
Each coef\/f\/icient $\psi_{i_1, \dots, i_N} $ is a~polynomial function of the parameters $a_{rst}$ in the $D \times D
\times d$ tensor~$A$.
Thus~\eqref{eqn-phi} def\/ines a~regular map $\Psi: \mathbb{C}^{D^2d} \to \mathbb{C}^{d^N}$, whose image we denote~by
$\PB(D,d,N)$, the set of tensors representable by translation-invariant matrix product states with
periodic boundary conditions.
Its closure $\overline\PB(D,d,N)$ in either the Zariski or classical topology is an irreducible algebraic
variety consisting of those tensors which can be approximated {\it arbitrarily well} by MPS.
We can thus ref\/ine Question~\ref{q:characterize} as follows.

\begin{question}
Fixing,~$D$,~$d$, and~$N$, what polynomial relations must the coefficients of a~matrix product state satisfy: what is
the defining ideal of $\overline\PB(D,d,N)$?
\end{question}

\begin{figure}[th]
\centering
\includegraphics{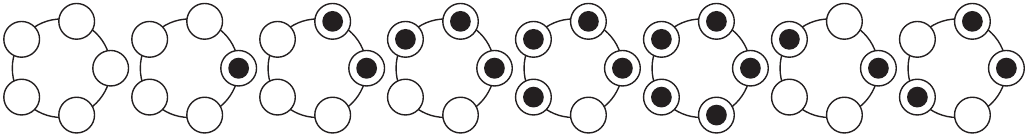}

\caption{The eight binary necklaces for $N=5$.}\label{Fig2}
\end{figure}

We primarily examine the fully binary case $D = d = 2$.
The invariance of trace under cyclic permutations of the matrices $A_{i_1}, \ldots, A_{i_N}$ means we can immediately
restrict to the subspace spanned by {\em binary necklaces} (equivalence classes of binary strings under cyclic
permuta\-tion)~\cite{OEIS}.
For $N=3$ physical legs, this is the coordinate subspace $(\psi_{000}:\psi_{100}:\psi_{110}:\psi_{111})$ and all
three-qubit states with cyclic symmetry are matrix product states.
For $N=4$ it is the six-dimensional coordinate subspace
$(\psi_{0000}:\psi_{1000}:\psi_{1100}:\psi_{1010}:\psi_{1110}:\psi_{1111})$ and not all states are MPS
(Theorem~\ref{thm-hyp}).
In the $N=5$ case the 8 equivalence classes of coef\/f\/icients under cyclic permutation are $\psi_{00000}$, $\psi_{10000}$,
$\psi_{11000}$, $\psi_{11100}$, $\psi_{11110}$, $\psi_{11111}$, $\psi_{10100}$, and $\psi_{11010}$ (see Fig.~\ref{Fig2}).
For $N=6,\dots, 15$ the dimensions of this ``necklace space'' are $14$, $20$, $36$, $60$, $108$, $188$, $352$, $632$,
$1182$, and $2192$~\cite{OEIS}.
In general there are
\begin{gather*}
n_d(N)=\frac{1}{N}\sum\limits_{\ell \mid N}\varphi(\ell)d^{N/\ell}
\end{gather*}
$d$-ary necklaces of length~$N$, where~$\varphi$ is Euler's totient function.
Thus translation invariant MPS with periodic boundary of length~$N$ and physical bond dimension~$d$ live in a~linear
space isomorphic to $\mathbb{C}^{n_d(N)}$.

Na\"ively we have 8 parameters in our $2   \times   2   \times   2$ tensor~$A$, but on each virtual bond we can apply
a~gauge transformation $P(\, \cdot \,)P^{-1}$ for $P \in {\rm SL}_2$ without changing the state~\cite{perez2007matrix}.
Since ${\rm SL}_2$ is $3$-dimensional, we expect $\overline\PB(2,2,3)$ to be 5-dimensional.
Counting this way, our {\em expected dimension} of $\overline\PB(D,d,N)$ is $\min\{D^2(d-1)+1, n_d(N)\}$.
We expect $\overline\PB(D,d,N)$ to be a~hypersurface when this equals $n_d(N)$, which happens f\/irst when
$(D,d,N)=(2,2,4)$.
In this case our expectation holds:

\begin{Theorem}
\label{thm-hyp}
A~four-qubit state~$\Psi$ is a~limit of binary periodic translation invariant MPS with $N=4$ if and only if the
following irreducible polynomial vanishes:
\begin{gather*}
\psi_{1010}^2\psi_{1100}^4-2\psi_{1100}^6-8\psi_{1000}\psi_{1010}\psi_{1100}^3\psi_{1110} +
12\psi_{1000}\psi_{1100}^4\psi_{1110}-
4\psi_{1000}^2\psi_{1010}^2\psi_{1110}^2
\\
\qquad {} +
2\psi_{0000}\psi_{1010}^3\psi_{1110}^2+16\psi_{1000}^2\psi_{1010}\psi_{1100}\psi_{1110}^2-
4\psi_{0000}\psi_{1010}^2\psi_{1100}\psi_{1110}^2
\\
\qquad {}-16\psi_{1000}^2\psi_{1100}^2\psi_{1110}^2 +
4\psi_{0000}\psi_{1010}\psi_{1100}^2\psi_{1110}^2-
4\psi_{0000}\psi_{1100}^3\psi_{1110}^2
\\
\qquad {}-
4\psi_{0000}\psi_{1000}\psi_{1010}\psi_{1110}^3+8\psi_{0000}\psi_{1000}\psi_{1100}\psi_{1110}^3
-
\psi_{0000}^2\psi_{1110}^4+2\psi_{1000}^2\psi_{1010}^3\psi_{1111}
\\
\qquad {}-\psi_{0000}\psi_{1010}^4\psi_{1111} -
4\psi_{1000}^2\psi_{1010}^2\psi_{1100}\psi_{1111}+
4\psi_{1000}^2\psi_{1010}\psi_{1100}^2\psi_{1111}
\\
\qquad {}
+2\psi_{0000}\psi_{1010}^2\psi_{1100}^2\psi_{1111}-4\psi_{1000}^2\psi_{1100}^3\psi_{1111}+
\psi_{0000}\psi_{1100}^4\psi_{1111}
\\
\qquad {} -4\psi_{1000}^3\psi_{1010}\psi_{1110}\psi_{1111}+
4\psi_{0000}\psi_{1000}\psi_{1010}^2\psi_{1110}\psi_{1111}+
8\psi_{1000}^3\psi_{1100}\psi_{1110}\psi_{1111}
\\
\qquad {} - 8\psi_{0000}\psi_{1000}\psi_{1010}\psi_{1100}\psi_{1110}\psi_{1111}-
2\psi_{0000}\psi_{1000}^2\psi_{1110}^2\psi_{1111}+2\psi_{0000}^2\psi_{1010}\psi_{1110}^2\psi_{1111}
\\
\qquad {}-
\psi_{1000}^4\psi_{1111}^2
+
2\psi_{0000}\psi_{1000}^2\psi_{1010}\psi_{1111}^2-\psi_{0000}^2\psi_{1010}^2\psi_{1111}^2.
\end{gather*}
\end{Theorem}

The proof of this theorem appears immediately after the proof of Proposition~\ref{proposition11}.
Hence, up to closure, the set $\PB(2,2,4)$ of tensors that can be represented in the
form~\eqref{eqn-phi}
where $A_0$ and $A_1$ are arbitrary $2\times 2$ matrices, is a~sextic hypersurface in the space of $2\times 2 \times 2
\times 2$ tensors invariant under cyclic permutations of the indices.
The 30-term hypersurface equation was found using a~parametrization of the matrices that is similar to the birational
parametrization of binary hidden Markov models given in~\cite{critch2013binary}.

An example of a~pure state on four qubits on which the polynomial~$f$ of Theorem~\ref{thm-hyp} is nonvanishing, and so
cannot be arbitrarily well approximated by such a~matrix product state, is given by letting
$\psi_{1010}=\psi_{1110}=-1/4$ and $\psi_{0000}=\psi_{1000}=\psi_{1100}=\psi_{1111}=1/4$.
In this example, $f(\Psi)=2^{-5}$, which is the maximal value of $f(\Psi)$ attained on corners of the $6$-D hypercube.

The other cases with $N\leq 15$ when we expect $\overline\PB$ to be a~hypersurface are when $(D,d,N)=$
$(2,4,6)$, $(3,3,7)$, $(5,15,12)$, $(3,71,13)$, and $(2,296,14)$.
In general, we will need many more polynomials to def\/ine the space of matrix product states as their zero locus.
As an example, consider $\overline\PB(2,2,5)$, which we expect to be a~f\/ive-dimensional variety in the
necklace space $\mathbb{C}^{8}=\mathbb{C}^{N_2(5)}$.

\begin{Theorem}\label{thm225}
Any homogeneous minimal generating set for the ideal of $\overline\PB(2,2,5)$ must contain exactly~$3$
quartic and $27$ sextic polynomials, possibly some higher degree polynomials, but none of degree~$1$,~$2$,~$3$, or~$5$.
\end{Theorem}

\begin{proof}
Using the bi-grading of Proposition~\ref{bihom}, we decompose the ideal~$I$ into vector spaces $I_{r,s}$.
For each $(r,s)$ with $\frac{1}{5}(r+s)\leq 6$, we select a~large number of parameter values $\widehat{A}$ at random,
and use Gaussian elimination to compute a~basis for the vector space $\widehat{I_{r,s}}$ of polynomials vanishing at
their images $\Psi(\widehat{A})$, which is certain to contain $I_{r,s}$.
We then substitute indeterminate entries for~$A$ symbolically into the polynomials to ensure that they lie in $I_{r,s}$,
yielding a~bihomogeneous basis for~$I$ in total degree $\leq 6$.
\end{proof}

This is interesting, because the variety only has codimension~$3$, but requires {\em at least} 30 equations to cut it
out ideal-theoretically.
Such a~collection of 3 quartics and 27 quadrics was found and verif\/ied symbolically.
Exact numerical tests (intersection with random hyperplanes) indicate that the top dimensional component of the ideal
they generate is reduced and irreducible of dimension~$5$, and is therefore equal to
$\overline\PB(2,2,5)$.

A detailed account of the computational commutative algebra and algebraic geometry methods needed to extend such results would take us too far af\/ield; we refer the interested reader to the textbooks \cite{Eisenbud2002,Greuel2008}.

\subsection[Homogeneity and ${\rm GL}_d$-invariance]{Homogeneity and $\boldsymbol{{\rm GL}_d}$-invariance}

Note that the equation of Theorem~\ref{thm-hyp} is homogeneous of degree~$6$, and every monomial has the same total number
of $1$'s appearing in its subscripts.
Every MPS variety will be homogeneous in such a~grading:

\begin{Proposition}
\label{bihom}
For any $D$, $d$, $N$, the space of translation-invariant MPS limits with periodic boundary conditions is cut out~by
polynomials in which each monomial has the same total number of $0$'s, $1$'s, $\dots$, $(d-1)$'s appearing in its subscripts.
\end{Proposition}

\begin{proof}
In fact we claim that the ideal of $\overline\PB(D,d,N)$ is $\mathbb{Z}^d$-homogeneous with respect to~$d$
dif\/ferent $\mathbb{Z}$-gradings $\deg_i$ for $0\leq i \leq d-1$, where $\deg_i(\Psi_J)$ is the number of occurrences
of~$i$ in~$J$.
Since $\deg(\psi_J):=\frac{1}{N}\sum\limits_{i=0}^{N-1}\deg_i(\psi_J)=1$, $\overline\PB(D,d,N)$ is also
homogeneous in the standard grading.

The usual parametrization~$\Psi$, where $A_0,\ldots,A_{d-1}$ have generic entries, is $\mathbb{Z}^d$-homogeneous with
respect to the grading above along with letting $\deg_i(A_j)=1$ when $i=j$ and $0$ when $i\neq j$.
Since $\Psi$ is a homogeneous map (as can be seen by writing out its coordinates), its kernel, the def\/ining ideal of $\overline\PB(D,d,N)$, is homogeneous in each of these gradings as well.
\end{proof}

In fact, the variety is homogeneous in a~stronger sense because of an action of ${\rm GL}_{d}$ on the parameter space
of~$\Psi$.
In the example above, the action is given~by
\begin{gather*}
\begin{pmatrix}g_{00}&g_{01}\\ g_{10}&g_{11}   \end{pmatrix}
\begin{pmatrix} A_0 \\ A_1\end{pmatrix} =  \begin{pmatrix} g_{00}A_0+g_{01}A_1 \\ g_{10} A_1+g_{11} A_1  \end{pmatrix},
\end{gather*}
which descends to an action on~$\Psi$~by
\begin{gather*}
\begin{pmatrix}g_{00}&g_{01}\\ g_{10}&g_{11}  \end{pmatrix}
\cdot \psi_{ijkl} = \sum_{pqrs} g_{ip}g_{jq}{g_{kr}}g_{ls}\psi_{pqrs}.
\end{gather*}

The embedding $(\mathbb{C}^*)^d\subset {\rm GL}_d$ as diagonal matrices gives rise to the $\mathbb{Z}^d$ homogeneity of the
proposition above.

\subsection{Linear invariants and ref\/lection symmetry}

There are additional symmetries peculiar to the case $D=d=2$.
For a~generic pair of $2\times 2$ matrices $A_0$, $A_1$, there is a~one-dimensional family of matrices $P\in {\rm SL}_2$ such
that $P^{-1}A_iP$ are symmetric.
Thus, a~generic point $\Psi\in \PB(2,2,N)$ can be written as $\Psi(A_0,A_1)$ with $A_i^T=A_i$, and then $
\Psi_J = \operatorname{tr} \big(\prod\limits_{j \in J} A_j\big) = \operatorname{tr} \big(\big(\prod\limits_{j \in J} A_j\big)^T\big) =
\operatorname{tr}\big(\prod\limits_{j \in \textnormal{reverse}(J)} A_j\big) = \Psi_{\textnormal{reverse}(J)}$.
This implies

\begin{Proposition}
If an~$N$-qubit state~$\Psi$ is a~limit of binary periodic translation invariant matrix product states, then it has
reflection symmetry: $\psi_J = \psi_{\textnormal{reverse}(J)}$ for all~$J$.
\end{Proposition}

For $N\geq 6$,~$N$-bit strings can be equivalent under ref\/lection but not cyclic permutation, so then
$\PB(2,2,N)$ admits additional linear invariants, i.e.\ linear polynomials vanishing on the model.
For $N = 6,7$ these are
\begin{gather*}
\PB(2,2,6):
\
  \psi_{110100}-\psi_{110010},
\\
\PB(2,2,7):
\
\psi_{1110100}-\psi_{1110010}
\qquad
\textnormal{and}
\qquad
\psi_{1101000}-\psi_{1100010}.
\end{gather*}

For small~$N$ we can f\/ind {\em all} the linear invariants of $\PB(2,2,N)$ using the bigrading of
Proposition~\ref{bihom} as in the proof of Theorem~\ref{thm225}.
Modulo the cyclic and ref\/lection invariants, there are no further linear invariants for $N\leq 7$, but
$\PB(2,2,8)$ has a~single ``non-trivial'' linear invariant
\begin{gather*}
\psi_{11010010}+\psi_{11001100}-\psi_{11001010} + \psi_{11101000}-\psi_{11011000}-\psi_{11100100},
\end{gather*}
which was obtained by direct calculation.
For $N=9$, $10$, $11$, and $12$, $\PB(2,2,N)$ admits~$6$, $17$, $44$, and $106$ such non-trivial
invariants, in each case unique up to change of basis on the vector space they generate.

\section{MPS with open boundary conditions}
\label{sec-bmps}

We now consider matrix product states with open boundary conditions, which are even more similar to hidden Markov models
than the periodic version.
Here the state is determined by two boundary state vectors $b_0,b_1\in\mathbb{C}^D$, along with the $D \times D$
parameter matrices $A_{0}, \dots, A_{d-1}$ of the MPS,~by
\begin{gather}
\label{eqn-phi2}
\Psi  = \sum\limits_{i_1, \dots, i_N} b_0^TA_{i_1} \cdots A_{i_N}b_1|i_1 i_2 \dots i_N \rangle
\\
\phantom{\Psi}
 = \sum\limits_{i_1, \dots, i_N} \operatorname{tr}(BA_{i_1} \cdots A_{i_N})|i_1 i_2 \dots i_N \rangle,
\label{eqn-phi2b}
\end{gather}
where $B=b_1b_0^T$ is a~rank 1 matrix.
We denote the set of states obtainable in this way by $\OB(D,d,N)$, and its closure (Zariski or classical)
by $\overline\OB(D,d,N)$.
We do not have the cyclic symmetries of the $\PB$ model here, so we consider
$\overline\OB(D,d,N)$ as a~subvariety of $\mathbb{C}^{d^N}$.
If the $A_i$ and $b_0^T$ have non-negative entries with row sums equal to~$1$, and $b_1$ is a~vector of $1$'s,
then~\eqref{eqn-phi2} is exactly the Baum formula for HMM, so in fact the model $\operatorname{HMM}(D,d,N)$ studied
in~\cite{critch2013binary} is contained in $\OB(D,d,N)$.

The expression~\eqref{eqn-phi2b} for~$\Psi$ is invariant under the action of ${\rm SL}_D$ on the $A_i$ and~$B$ by simultaneous
conjugation.
Thus, we may assume~$B$ is in Jordan normal form, i.e.~a matrix of all zeroes except possibly in the top left corner.
As well, the map $(B,A_1,\ldots,A_d)\mapsto(t^{-N}B,tA_1,\ldots,tA_d)$ preserves~$\Psi$, so discarding the case $B=0$
(which will not change $\overline\OB$) we can assume that the top left entry of~$B$ is 1.
Thus~$\Psi$ is determined by $dD^2$ parameters, the entries of the $A_i$.
In particular, $\overline\OB(2,2,3)$ is parametrized by (a dominant map from) $8$ parameters, and lives in
an $8$-dimensional space.
This parametrization still turns out still to be degenerate:

\begin{Theorem}\label{thm-hyp2}
A~three-qubit state~$\Psi$ is a~limit of $N=3$ binary translation invariant MPS with open boundary conditions if and
only if the following $22$-term quartic polynomial vanishes:
\begin{gather*}
\psi_{011}^2\psi_{100}^2 - \psi_{001}\psi_{011}\psi_{100}\psi_{101} - \psi_{010}\psi_{011}\psi_{100}\psi_{101} +
\psi_{000}\psi_{011}\psi_{101}^2+
\psi_{001}\psi_{010}\psi_{011}\psi_{110}
\\
\qquad {} - \psi_{000}\psi_{011}^2\psi_{110} - \psi_{010}\psi_{011}\psi_{100}\psi_{110}+
\psi_{001}\psi_{010}\psi_{101}\psi_{110} + \psi_{001}\psi_{100}\psi_{101}\psi_{110}
\\
\qquad {} - \psi_{000}\psi_{101}^2\psi_{110} -
\psi_{001}^2\psi_{110}^2+
\psi_{000}\psi_{011}\psi_{110}^2 - \psi_{001}\psi_{010}^2\psi_{111} + \psi_{000}\psi_{010}\psi_{011}\psi_{111}
\\
\qquad {} +
\psi_{001}^2\psi_{100}\psi_{111}+
\psi_{010}^2\psi_{100}\psi_{111} - \psi_{000}\psi_{011}\psi_{100}\psi_{111} - \psi_{001}\psi_{100}^2\psi_{111}
\\
\qquad {} -
\psi_{000}\psi_{001}\psi_{101}\psi_{111}
+
\psi_{000}\psi_{100}\psi_{101}\psi_{111} + \psi_{000}\psi_{001}\psi_{110}\psi_{111} -
\psi_{000}\psi_{010}\psi_{110}\psi_{111}.
\end{gather*}
That is, the variety $\overline\OB(2,2,3)$ is a~quartic hypersurface in $\mathbb{C}^8$ cut out by the
polynomial above.
This polynomial previously appeared in the context of the HMM~{\rm \cite{pachter2004tropical}}.
\end{Theorem}

\begin{proof}\looseness=-1
The map~$\Psi$ and its image are homogeneous in the same grading as described in Proposition~\ref{bihom}, which we can use as in
the proof of Theorem~\ref{thm225} to search for low degree polynomials vanishing on the variety.
When $(D,d,N)=(2,2,3)$ the quartic from the theorem appears in this search.
The quartic is prime, and therefore def\/ines a~$7$-dimensional irreducible hypersurface in $\mathbb{C}^8$.
On the other hand, the Jacobian of the map~$\Psi$ at a~random point, e.g.\
the point where $A_0$, $A_1$ have entries $1$, $2$, $3$, $4$, $5$, $6$, $7$, $8$ in that order, has rank~$7$.
Therefore $\overline\OB(2,2,3)$ is of dimension at least~$7$, and contained in the quartic hypersurface
above, so they must be equal.
\end{proof}

From Theorem~\ref{thm-hyp2}, we can derive conditions on $\OB(2,2,N)$ for $N\geq 4$ as well.
There is an~{\em improper marginalization map} from $\OB(2,2,N)$ to $\OB(2,2,3)$ given~by
$\Psi_{I}\mapsto \sum\limits_{|J|=N-3} \Psi_{IJ}$ for each~$I$ of length~$3$, which commutes with the assignment
$b_1\mapsto \sum\limits_{j_3,\ldots, j_N} A_{j_3}\cdots A_{j_N} b_1$.
In fact there are $N-2$ such improper marginalization maps, each given by choosing $3$ consecutive indices~$I$ to marginalize to
(summing over the remaining indices~$J$).
By composing these maps with the quartic polynomial above, we can obtain $N-2$ quartic polynomials vanishing on
$\OB(2,2,N)$. Note that this improper marginalization is not the quantum marginal obtained by a~partial trace of the density operator.  There are experimental methods to improperly marginalize a~MPS, e.g.\ by postselection on the summed-over indices.

By analogy to the case of hidden Markov models discussed in the next section, we make the following

\begin{Conjecture}
For $N\geq 4$, a~generic~$N$-qubit state can be recovered from its improper margina\-li\-zation to any three consecutive states.
That is, each improper marginalization map $\overline\OB(2,2,N)\to\overline\OB(2,2,3)$ is
a~birational equivalence of varieties.
\end{Conjecture}

The analogous statement with $\overline{\operatorname{HMM}}$ in place of $\overline\OB$ is shown
to be true in~\cite{critch2013binary}.

Related results include \cite{Erickson1970} and \cite{Vidyasagar2011}, where it is shown that a quasi-realization for a HMM can be obtained from moments of order~$2k+1$, where $k$ is the word length at which the matrix  $H_{uv} = [P(u^*v), |u|=|v|=k]$ achieves rank~$r$.

Although the notion of quantum marginalization is very dif\/ferent from classical marginalization, from the point of view of algebraic geometry the loss of information about which point on the variety we began with may not be signif\/icant.  A more natural conjecture which would have direct relevance for quantum information is the following.

\begin{Conjecture}
A~generic~$N$-qubit $(D=d=2)$ translation invariant matrix product state~$\Psi$ with open boundary conditions is
determined up to phase by a~reduced density operator which traces out all but a~chain of three adjacent states, but no
fewer.
\end{Conjecture}

Such results would be useful for quantum state tomography when tensor network state assumptions hold. When the three adjacent states are qubits~$1$,~$2$, and $3$ (the f\/irst three legs of the diagram), this amounts to
saying that the group $S^1$ of unit-modulus complex numbers acts transitively on generic f\/ibres of the real-algebraic
map
\begin{gather*}
\Psi \mapsto \left(\sum\limits_{i_4,\ldots,i_N}\Psi_{j_1j_2j_3i_4\ldots i_N}\Psi^\dagger_{k_1k_2k_3i_4\ldots
i_N}\right)_{j_1j_2j_3k_1k_2k_3}
\end{gather*}
when restricted to $\OB(2,2,N)$.
Here the right hand side denotes an order 6 tensor with indices $j_1$, $j_2$, $j_3$, $k_1$, $k_2$, $k_3$, and
$\Psi^\dagger$ denotes complex conjugation.

\begin{Conjecture}\label{conjecture10}
A~generic~$N$-qubit $(D=d=2)$ periodic translation invariant matrix product state~$\Psi$ is determined up to phase~by
a~reduced density operator which traces out all but a~chain of four adjacent states, but no fewer.
\end{Conjecture}

Again, although the classical and quantum marginals are very dif\/ferent, from the point of view of algebraic geometry there is reason to hope that if one provides suf\/f\/icient information about which point on the MPS variety we began with, so will the other.
Similarly Conjecture~\ref{conjecture10} amounts to saying that $S^1$ acts transitively on generic f\/ibres of the map
\begin{gather*}
\Psi \mapsto
\left(\sum\limits_{i_5,\ldots,i_N}\Psi_{j_1j_2j_3j_4i_5\ldots i_N}\Psi^\dagger_{k_1k_2k_3k_4i_5\ldots i_N}\right)_{j_1j_2j_3j_4k_1k_2k_3k_4}
\end{gather*}
when restricted to $\PB(2,2,N)$.

\section{Matrix product states as complex valued hidden\\ Markov models}
\label{sec:HMM}

We now explain how the polynomial in Theorem~\ref{thm-hyp} was obtained, and connect the classical {\em hidden Markov
model} and matrix product states through a~reparametrizing rational map.
The parametrization of the state~$\Psi$ is analogous to that of the moment tensor of a~binary hidden Markov model used
in~\cite{critch2013binary} for symbolic computations.

\begin{figure}[t]
\centering
\includegraphics[scale=0.95]{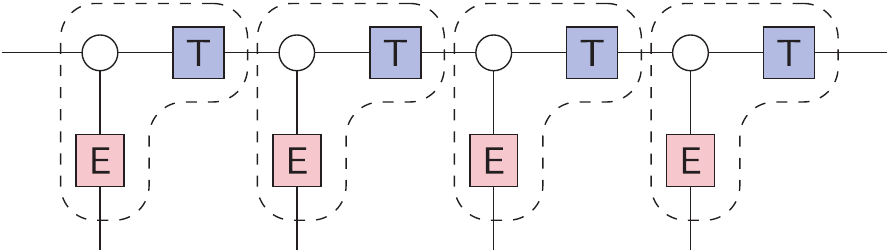}

\caption{Parameterization of an MPS model as a~complex HMM using complex~$E$ and~$T$ matrices with all row sums equal to
$z \in \mathbb{C}$ and copy dot (comultiplication) tensor (circle).
Contraction of a~region of the tensor network enclosed by a~dashed line yields an~$A$ tensor.}
\label{fig:HMMparam}
\end{figure}

\looseness=-1
The fact that MPS can be seen as quantum analogues of HMMs is well known in quantum probability.  Here we show that this connection is more than an analogy, by giving an explicit HMM-motivated parametrization of an MPS~$\Psi$ which specializes to an HMM probability distribution in the case where all the parameters are real stochastic matrices.   While from the quantum probability perspective it is often the density matrix rather than the MP vector state that plays the role of the probability distribution of the HMM, note that here the analogy is between vector state and probability distribution.  This relationship is useful because it makes some of the algebraic results from the classical case applicable, because it removes internal symmetries in a~natural way, and because it provides a means to generalize classical statistical results (and algorithms) to the quantum case whenever such maps can be written down.  The map between HMM and MPS we describe can be compactly expressed in the language of string diagrams as shown in Fig.~\ref{fig:HMMparam}.

Let~$T$ be a~$2 \times 2$ {\em transition} matrix and~$E$ a~$2 \times 2$ {\em emission} matrix.
For a~(classical) hidden Markov model,~$T$ and~$E$ are nonnegative stochastic matrices (their rows sum to one),
repre\-sen\-ting a~four-dimensional parameter space.
For $\PB$,~$T$ and~$E$ will be complex with row sums all equal to some constant $z \in \mathbb{C}$, so
they form a~parameter space isomorphic to $\mathbb{C}^5$.
Note that from the standpoint of projective geometry, exchanging a~requirement that rows sum to one to a~requirement that they sum to shared,
arbitrary complex number is actually natural.  This is a~somewhat common
trick in algebraic statistics when working with computer algebra systems.
We parametrize the $A_i$ in terms of $(T,E)$~by
\begin{gather*}
A_0 = T,
\qquad
A_1 =
\begin{pmatrix}
e_{01}& 0 \cr 0 & e_{11}
\end{pmatrix}
\begin{pmatrix}
t_{00}& t_{01} \cr t_{10} & t_{11}
\end{pmatrix}.
\end{gather*}
This is shown in Fig.~\ref{fig:HMMparam}; grouping and contracting the~$E$,~$T$, and copy dot tensors into an~$A$
tensor yields a~dense parameterization of an MPS as depicted in Fig.~\ref{fig:TIMPSPB}.
We then parameterize~$E$ and~$T$ with the f\/ive parameters $u$, $v_0$, $b$, $c_0$, $z$ by setting
\begin{gather*}
E =
\begin{pmatrix}
z-u+v_0& u-v_0 \cr z-u-v_0 & u+v_0
\end{pmatrix}
\qquad
\text{and}
\qquad
T =
\begin{pmatrix}
z+u-c_0& z-b+c_0 \cr z-b-c_0 & z+b+c_0
\end{pmatrix}.
\end{gather*}
Composing these formulae with the map $(A_0,A_1)\mapsto \Psi$ yields a~restricted parametrization $\rho_N: \mathbb{C}^5
\to \mathbb{C}^{2^N}$, whose image lies inside $\PB(2,2,N)$.

\begin{Proposition}\label{proposition11}
The variety $\overline\PB(2,2,N)$ is at most $5$-dimensional, and the image of our restricted
parametrization $\rho_N$ is dense in it.
\end{Proposition}
\begin{proof}
Suppose $\Psi = \Psi(A_0,A_1)$ for $A_0$, $A_1$ generic.
First, we will transform the $A_i$ by simultaneous conjugation with an element~$P$ of ${\rm SL}_2$ to a~new pair of matrices
$A_0'$, $A_1'$ such that $A_0'$ has equal row sums and $A_1'=DA_0'$ for a~diagonal matrix~$D$.
Generically, $A_0$ is invertible, and we can diagonalize the matrix $A_1A_0^{-1}$, so we write $U^{-1}A_1A_0^{-1}U=D_0$,
and then $U^{-1}A_1U=D_0U^{-1}A_0U$.
Next we f\/ind another diagonal matrix $D_1\in {\rm SL}_2$ such that $D_1^{-1}U^{-1}A_0UD_1$ has equal row sums.
Then let $P=UD_1$ and $A_i'=D_1^{-1}U^{-1}A_iUD_1$, and we are done with our transformation.
Now $\Psi = \Psi(A_0',A_1')$ since simultaneous conjugation does not change trace products.
But now letting~$z$ be the common row sum of $A_0'$, we can solve linearly for~$u$, $v_0$,~$b$, and $c_0$ to obtain
$\Psi = \rho(u,v_0,b,c_0,z)$.
\end{proof}

In fact we know from exact computations in Macaulay2~\cite{M2} that the dimension $\dim \PB(2,2,N)$ $= 5$ for
$4\leq N \leq 100$.
This is proven by checking that the Jacobian of~$\rho$ attains rank $5$ at some point with randomly chosen integer
coordinates, giving a~lower bound of $5$ on the dimension of its image. We can now prove Theorem~\ref{thm-hyp}.

\begin{proof}[Proof of Theorem~\ref{thm-hyp}]
When parametrized using~$\rho$, there are suf\/f\/iciently few parameters and the entries of~$\Psi$ are suf\/f\/iciently short expressions that Macaulay2 is also able to compute the~exact kernel of the parametrization, i.e.\ def\/ining equations for the model.  It is by this method that we obtain the hypersurface equation of Theorem~\ref{thm-hyp} as the only ideal generator for $\operatorname{PB}(2,2,4)$.
\end{proof}

\subsection{Identifying parameters of MPS}

Determining the parameters of an MPS is related to {\em quantum state tomography}, and represents a~quantum analog to
the identif\/iability problem in statistics.
The extent to which the parameters can be identif\/ied can be addressed algebraically.

Given $D\times D$ matrices $A_1,\ldots, A_d$ with indeterminate entries, we write $\mathcal{C}_{D,d}$ for the algebra of
polynomial expressions in their entries that are invariant under simultaneous conjugation of the matrices by ${\rm GL}_2$.

Sibirskii~\cite{sibirskii1968algebraic}, Leron~\cite{leron1976trace}, and Procesi~\cite{procesi1976invariant} showed that
the algebra $\mathcal{C}_{D,d}$ is generated by the traces of products $\operatorname{tr}(A_{i_0} \cdots A_{i_n})$ as $n\geq 0$ varies.
For this reason, $\mathcal{C}_{D,d}$ is called a~{\em trace algebra}.
Its spectrum, $\operatorname{Spec} \mathcal{C}_{D,d}$, is a~{\em trace variety}.
Since the coordinate ring of $\overline\PB(D,d,N)$ is a~subring of~$\mathcal{C}_{D,d}$, we have a~map
$\operatorname{Spec} \mathcal{C}_{D,d} \to \mathbb{C}^{d^N}$ parameterizing a~dense open subset of
$\overline\PB(D,d,N)$.

In the case $D=2$, Sibirskii showed further that the trace algebra $\mathcal{C}_{2,d}$ is minimally generated by the
elements $\operatorname{tr}(A_i)$ and $\operatorname{tr}(A_i^2)$ for $1\leq i \leq d$, $\operatorname{tr}(A_iA_j)$ for $1 \leq i < j \leq d$, and $\operatorname{tr}(A_iA_jA_k)$ for $1 \leq i < j < k \leq d$.

For $d=1,\ldots, 6$, the number of such generators is $2$, $5$, $10$, $18$, $30$, $47$.
In particular, when $d=2$, the number of generators equals the transcendence degree of the ring, $5=8-3$.
This means $\operatorname{Spec} C_{2,2}$ is isomorphic to~$\mathbb{C}^5$, yielding for each~$N$ a~dominant
parametrization $\phi_N: \mathbb{C}^5 \to \overline\PB(2,2,N)$.
Gr\"obner bases for randomly chosen f\/ibers indicate that for $N=4,\ldots, 10$, the map~$\phi_N$ is generically~$k$-to-one,
where $k=8,5,6,7,8,9,10$, respectively.
Continuing this sequence suggests the following.

\begin{Conjecture}
Using the trace parameterization $\phi_N$, for $N\geq 5$, almost every periodic boun\-dary MPS has exactly~$N$ choices of
parameters that yield it.
\end{Conjecture}
In other words, for $N\geq 5$, the parametrization $\phi_N: \mathbb{C}^5 \simeq \operatorname{Spec} C_{2,2} \to
\overline\PB(2,2,N)$ is generically~$N$-to-1.
Generically, the points of $\operatorname{Spec} C_{2,2}$ are in bijection with the ${\rm SL}_2$-orbits of the tensors~$A$.
The conjecture implies that, up to the action of~${\rm SL}_2$, the parameters of a~binary, $D=d=2$ translation invariant
matrix product state with periodic boundary are algebraically identif\/iable from its entries.

\section{Conclusion}
\label{sec:conc}

A~conjectured dictionary between tensor network state models and classical probabilistic graphi\-cal models was presented
in~\cite{Morton2012Benasque}.
In this dictionary, matrix product states correspond to hidden Markov models, the density matrix renormalization group
(DMRG) algorithm to the forward-backward algorithm, tree tensor networks to general Markov models, projected entangled
pair states (PEPS) to Markov or conditional random f\/ields, and the multi-scale entanglement renormalization ansatz
(MERA) loosely to deep belief networks.

In this work we formalize the f\/irst of these correspondences and use it to algebraically cha\-rac\-terize quantum states
representable by MPS and study their identif\/iability.
In future work we plan to extend these results to larger bond and physical dimensions, as well as to other tensor network state models such as tree tensor networks.  Some of these extensions should be straightforward, while others will require new ideas.

\subsection*{Acknowledgments}
AC and JM were supported in part by DARPA under awards FA8650-10-C-7020 and N66001-10-1-4040 respectively.
We would like to thank J.~Biamonte, J.~Eisert, B.~Sturmfels, F.~Vaccarino, F.~Verstraete, and G.~Vidal for helpful
discussions.
We are also grateful to anonymous referees who provided helpful comments and corrections.

\pdfbookmark[1]{References}{ref}
\LastPageEnding

\end{document}